\documentclass[aps,pre,amsmath,amssymb,twocolumn,groupedaddress,superscriptaddress,showpacs]{revtex4-1}

\usepackage{graphicx}
\usepackage{bm}

\usepackage{subfigure}
\begin{document}

\title[]{Demixing of active particles in the presence of external fields}

\author{Sunita Kumari}

\author{Andr\'e S. Nunes}

\author{Nuno A. M. Araujo}
 \email{nmaraujo@fc.ul.pt}

\author{Margarida M. Telo da Gama}

\affiliation{Departamento de Fisica, Faculdade de Ciencias, Universidade de
Lisboa, P-1749-016 Lisboa, Portugal and Centro de Fisica Teorica e
Computational, Universidade de Lisboa, P-1749-016, Lisboa, Portugal}

\begin{abstract}
Active systems are inherently out of equilibrium, as they collect energy from
their surroundings and transform it into directed motion. A recent theoretical
study suggests that binary mixtures of active particles with distinct effective
diffusion coefficients exhibit dynamical demixing above a threshold ratio of
the diffusion coefficients. Here, we show that this threshold may be reduced
drastically in the presence of external fields. We investigate the demixing as
a function of the ratio of the diffusion coefficients and discuss the
implications of the results for active systems.
\end{abstract}

\maketitle

\section{\label{sec:level1}Introduction}
Multicomponent systems may demix dynamic when the response of distinct species
to external fields differ
significantly~\cite{ahm73,dzu02,kud04,leu05,gcma05,reis06,vis11,riv11,riv12,grun16}.
Typical examples are mixtures of colloidal particles with different
electrophoretic mobilities in the presence of an external
field~\cite{leu05,vis11} or granular mixtures of particles with different
masses on vibrating plates~\cite{riv11,riva11}. However, dynamical demixing
does not necessarily require an external field. In active systems, demixing has
been observed for mixtures of particles with significantly different
activities~\cite{sten15} or effective diffusion coefficients~\cite{sim16}. 

Active particles are self-propelled random walkers that collect energy from
their surroundings and convert it into propelled motion~\cite{Ram10}. This
constant flow of energy, drives active particles away from thermodynamic
equilibrium leading to interesting new features that drive several active lines
of research~\cite{bech16}. One of the simplest theoretical models for the
dynamics of active particles consists of two-dimensional Brownian motion with
translational diffusion coefficient $D_\mathrm{T}$, and an additional
propulsion force of strength $v$, along a preferred direction. This direction
is coupled to the rotational degrees of freedom and evolves in time as a
Brownian process with rotational diffusion coefficient $D_\mathrm{R}$. The
overall dynamics will depend on the ratios of these values. While at very short
times, the motion is ballistic along the preferred direction, at significantly
long times, Brownian dynamics is observed with an effective translational
diffusion coefficient
$D_\mathrm{eff}=D_\mathrm{T}+\frac{1}{2}v^2/D_\mathrm{R}$~\cite{bech16}.

Living active systems often consist of mixtures of passive and active particles
or species with different
activities~\cite{tre09,ang11,chai11,dres11,sten15,nir14,kum15,taka15,sme17,yang14,gros15,tana17}.
Recently, Weber \textit{et al.} investigated the dynamics of binary mixtures of
active particles with different diffusion coefficients and the same Stokes
coefficient (mobility)~\cite{sim16}. They considered two species with diffusion
coefficients $D_\mathrm{L}$ and $D_\mathrm{H}$, respectively, with
$D_\mathrm{H}\geq D_\mathrm{L}$. They found that, at sufficiently low
$D=D_\mathrm{L}/D_\mathrm{H}$, the system demixes dynamically, forming
solid-like clusters of slow diffusing particles surrounded by a gas of rapidly
diffusing ones. For a packing fraction of $0.5$, demixing occurs at $D\lesssim
10^{-2}$ and this threshold decreases as the packing fraction decreases.

In what follows, we discuss the possibility of using external fields to promote
dynamical demixing. External stimuli (e.g., chemical/thermal gradients and
electromagnetic fields) modify the particles trajectories, through the creation
of distinct regions that localize one or both
species~\cite{adl66,arm03,blak75,ant13,erik16,boy16}. For simplicity, we
consider periodic (sinusoidal) potentials and identical responses to external
field, i.e., identical electrophoretic mobilities in external electromagnetic
fields. The results of the Brownian dynamics simulations, reported below,
reveal that as the amplitude of the external potential increases the threshold
value of $D$ increases. Also, the solid-like cluster of slow diffusing
particles is always localized along the minimum of the potential.

The paper is organized as follows. The description of the model and the
simulations are presented in the next Section. We report and discuss the
results in Sec.~\ref{sec:3}. Finally, we draw some conclusions in
Sec.~\ref{sec:4}.

\section{Model and Simulation\label{sec:2}}
\begin{figure*}[t]
\includegraphics[width=0.8\textwidth]{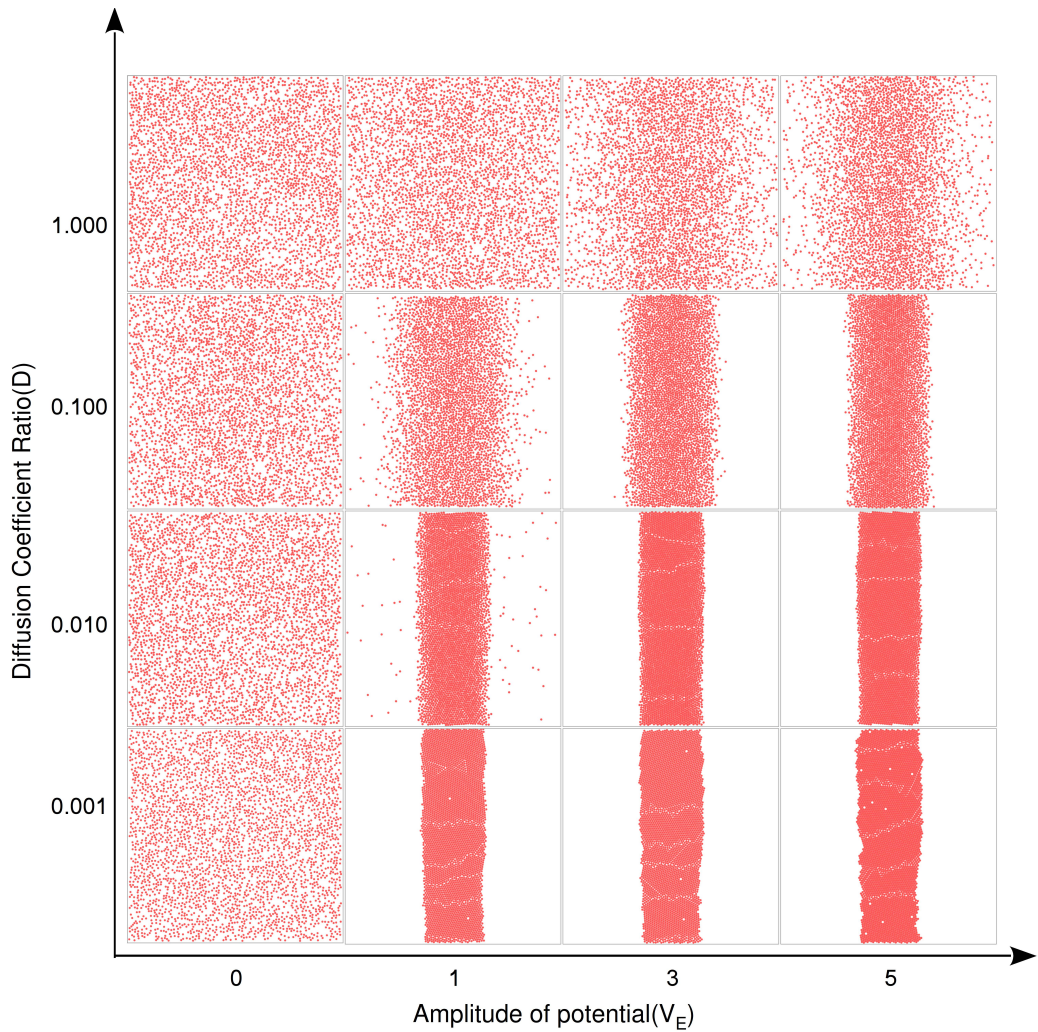}
\caption{Snapshots for single-component systems of particles with diffusion
	coefficient $D_\mathrm{L}$, with $N= N_{L} = 3184$ at $t = 10^{7}$, on
	a box of lateral dimension $L_b=200$. The average density is $\rho =
	0.5$.  Different rows are for different values of
	$D=D_\mathrm{L}/D_\mathrm{H}$ and different columns correspond to
	different values of the amplitude of the potential $V_{E}$. Note that,
	when changing $D$, we fixed the value of $D_\mathrm{H}$ and changed the
value of $D_\mathrm{L}$.~\label{fig::singlecomp}}
\end{figure*}
As in Ref.~\cite{sim16}, we consider a two-dimensional binary mixture of $N$
particles (disks). $N_\mathrm{H}=N/2$ particles are of species $\mathrm{H}$ and
the other $N_\mathrm{L}=N/2$ are of species $\mathrm{L}$, corresponding to
different effective diffusion coefficients $D_\mathrm{H}$ and $D_\mathrm{L}$,
respectively, where $D_\mathrm{H}\geq D_\mathrm{L}$. Particles move in a square
box of linear dimension $L_b$ with periodic boundary conditions in both
directions.  Since we are interested in timescales that are much larger than
the Stokes time, we will consider that all particles are in the overdamped
regime. Accordingly, the stochastic trajectory of particle $i$ is obtained by
solving the equation,
\begin{equation}\label{eq::dyn}
\dot{\vec{r}}_{i} =  \vec{\eta}_{i}(t) + \mu \sum_{j=1}^{N}\vec{F}_{ij}(t) -
\nabla_{\vec{r}_i}V_\mathrm{ext}(\vec{r}_{i}),
\end{equation}
where $\vec{r}_{i}$ is the position of the particle. Particle interactions are
given by short-ranged repulsive forces, $\vec{F}_{ij}=\alpha(2a -
r_{ij})\hat{r}_{ij}$, accounting for particle overlap ($r_{ij} < 2a$), where
$a$ is the particle radius, and vanishing $\vec{F}_{ij} = 0$, otherwise;
$r_{ij} = |\vec{r}_{i} - \vec{r}_{j}|$ is the distance between the particle
centers and $\hat{r}_{ij} = (\vec{r}_{i}-\vec{r}_{j})/r_{ij}$. $\alpha\geq 0$
is the strength of repulsion.  $\mu$ is the inverse of the Stokes coefficient
and it is the same for both species. $\vec{\eta}_{i}$ is a Gaussian random
white-noise term of zero mean ($\langle\eta_{i}\rangle = 0$) and correlation
$\langle\eta_{in}(t)\eta_{jl}(t')\rangle =
2D_{i}\delta_{ij}\delta_{nl}\delta(t-t')$, where $D_i$ is the particle
diffusion coefficient, $\delta_{ij}$ and $\delta_{nl}$ are Kronecker deltas,
where $n$ and $l$ are the spatial components of the vectors $\vec{\eta}_i$ and
$\vec{\eta}_j$, respectively, and $\delta(t-t')$ is the Dirac delta function.
Note that, since $\mu$ is the same for both species but $D_i$ is not, the
dynamics described by Eq.~(\ref{eq::dyn}) does not obey the fluctuation
dissipation theorem.  In fact, the dynamics describes a mixture of passive
particles at two distinct effective temperatures. The last term in
Eq.~(\ref{eq::dyn}) represents the interaction with the external potential. We
consider,
\begin{equation}\label{eq::potential}
V_\mathrm{ext}(x) = V_{E}\cos(kx) ,
\end{equation}
where $V_E$ is the amplitude of the potential. We take $k= \frac{2\pi}{L_{b}}$,
so that there is a single minimum at the center of the simulation box. Without
loss of generality, we parameterize the system in terms of the ratio
$D=D_\mathrm{L}/D_\mathrm{H}$, by fixing $D_\mathrm{H}$ and varying
$D_\mathrm{L}$ and we define the density as $\rho=N\pi a^2/L_b^2$.

We performed simulations for square boxes of linear dimensions $L_b =
\left\{60, 80, 100, 120, 150, 200, 600, 1000\right\}$, in units of the particle
radius $a$, for $\alpha= 90$ and $\mu=1$. We integrated Eq.~(\ref{eq::dyn})
using a second-order stochastic Runge-Kutta scheme~\cite{Branka}, with a
discrete time step $dt=2\times 10^{-3}$. Time is measured in units of
$a^2/D_\mathrm{H}$.

\section{Results\label{sec:3}}
\begin{figure}
\includegraphics[width=\columnwidth]{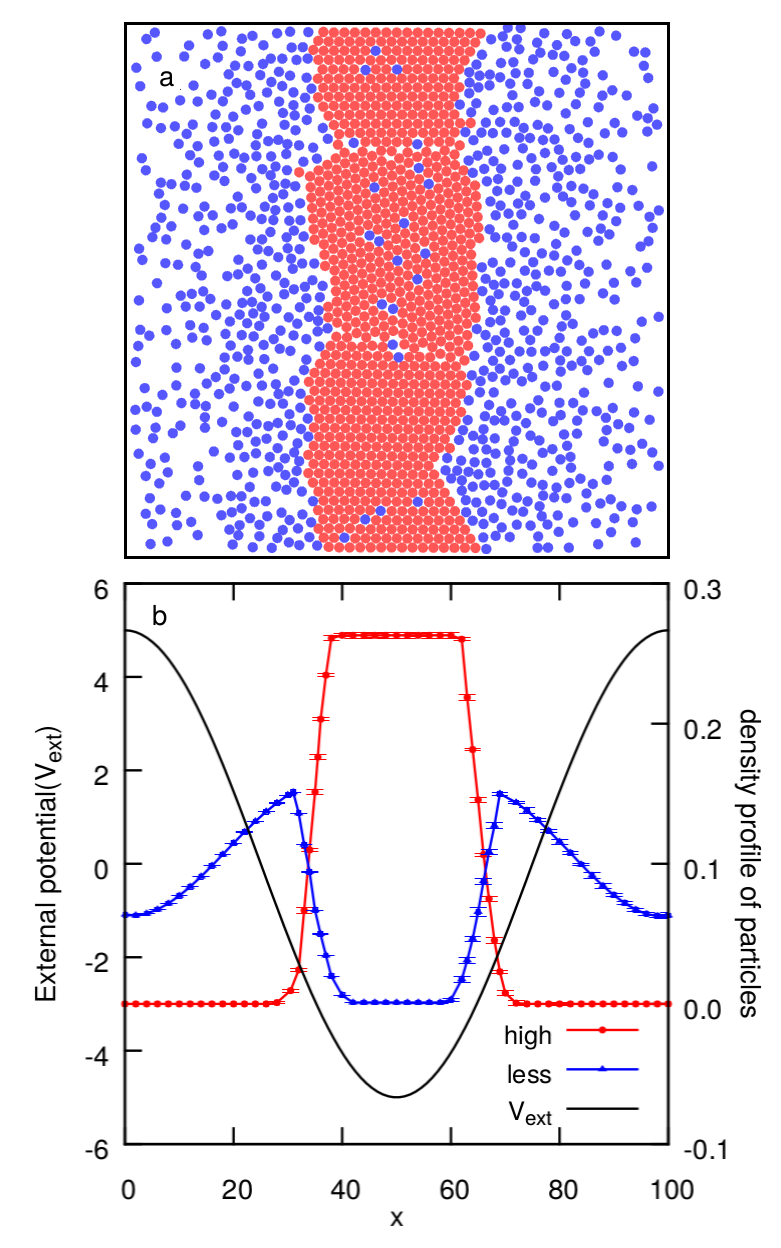}
\caption{(a) Snapshot and (b) density profile along the $x$-direction for a
	binary mixture of particles with low (red/light gray) and high
	(blue/dark gray) diffusion coefficients, for $N_{L} = N_{H} = 796$,
	$\rho = 0.5$, $D = 10^{-3}$, and $V_{E}=5$ at $t = 10^{7}$, in a box of
lateral dimension $L_b=100$. Each data point is an average over $25$ samples.
The error bars are smaller than the size of the symbols.
~\label{fig::binarymixg}}
\end{figure}
Let us start by considering a single component system.
Figure~\ref{fig::singlecomp} depicts snapshots for different values of $V_E$
and diffusion coefficient $D_\mathrm{L}$, at a constant density $\rho=0.5$,
with initial uniform distribution.  We use $D=D_\mathrm{L}/D_\mathrm{H}$, where
$D_\mathrm{H}=1$, everywhere in this work. As shown in the figure, in the
absence of the external potential ($V_E=0$), the particles are uniformly
distributed, but for $V_E\neq 0$, the particles accumulate around the minimum
of the potential (in the middle of the box) and a band is formed along the
$y$-direction.  The width of the band decreases as the amplitude of the
external potential increases while it increases as the diffusion coefficient
increases.  These results are in agreement with previous detailed studies of
static and dynamical properties of single-component passive systems in the
presence of similar external potentials~\cite{and16}.

We now consider equimolar binary mixtures ($N_\mathrm{L}=N_\mathrm{H}=796$), at
the same total density $\rho=0.5$, with a uniform initial distribution.
Figure~\ref{fig::binarymixg} depicts a snapshot (Fig.~\ref{fig::binarymixg}(a))
and the density profile (Fig.~\ref{fig::binarymixg}(b)) in the stationary state
($t=10^7$), for $V_E=5$ and $D=10^{-3}$.  The particles with low diffusion
coefficients (red particles) tend to accumulate in the center of the band, in
an ordered structure, surrounded by those with high diffusion coefficients
(blue particles). Note that, for the given values of $D_\mathrm{L}$ and
$D_\mathrm{H}$, both species are expected to form a central band in the
single-component system, as shown in Fig.~\ref{fig::singlecomp}, for
$D=10^{-3}$ and $D=1$, respectively.
\begin{figure*}
\includegraphics[width=0.8\textwidth]{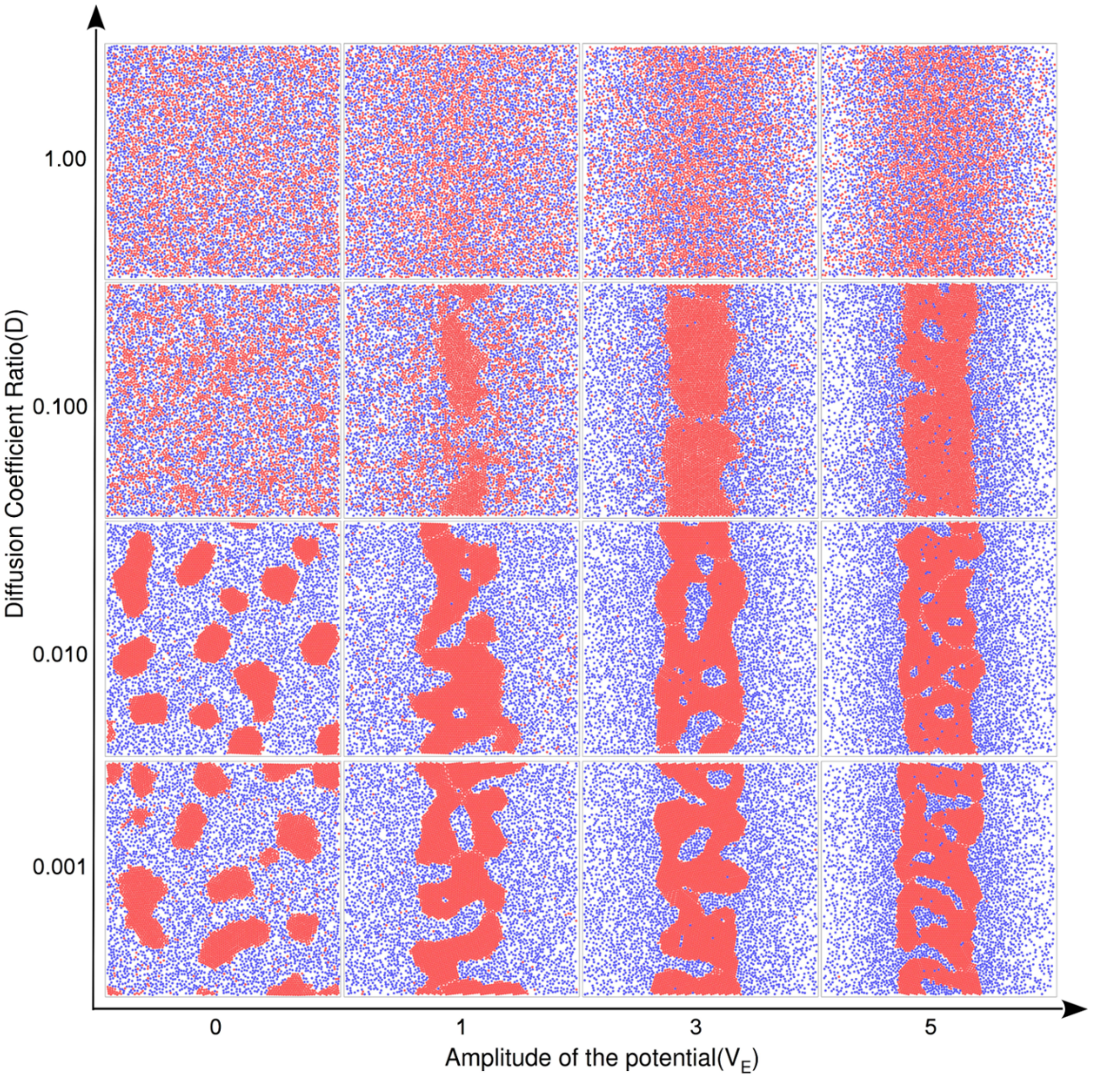}
\caption{Snapshots of binary mixtures, with $N_{L} = N_{H} = 3184$ particles
	with low (red/light gray) and high (blue/dark gray) diffusion
	coefficients at $t = 10^{7}$, in a box of lateral dimension $L_b=200$.
	The total density is $\rho = 0.5$.  Different rows depict different
values of $D$ and different columns different values of the external potential
amplitude $V_E$.~\label{fig::dandve}}
\end{figure*}

In order to analyze the dependence of the dynamics on $D$ and $V_E$, we
illustrate in Fig.~\ref{fig::dandve} snapshots of the stationary state at
different values of these parameters.  While in the absence of potential
($V_E=0$) demixing is observed only at low values of $D$ ($D<0.1$), when the
potential is switched on, $V_E\neq 0$, demixing occurs in the entire range of
$D$, for strong enough $V_E$. In all cases, the particles with the low
diffusion coefficient are those accumulating in the center of the band, forming
a very compact cluster.

\begin{figure}
\includegraphics[width=\columnwidth]{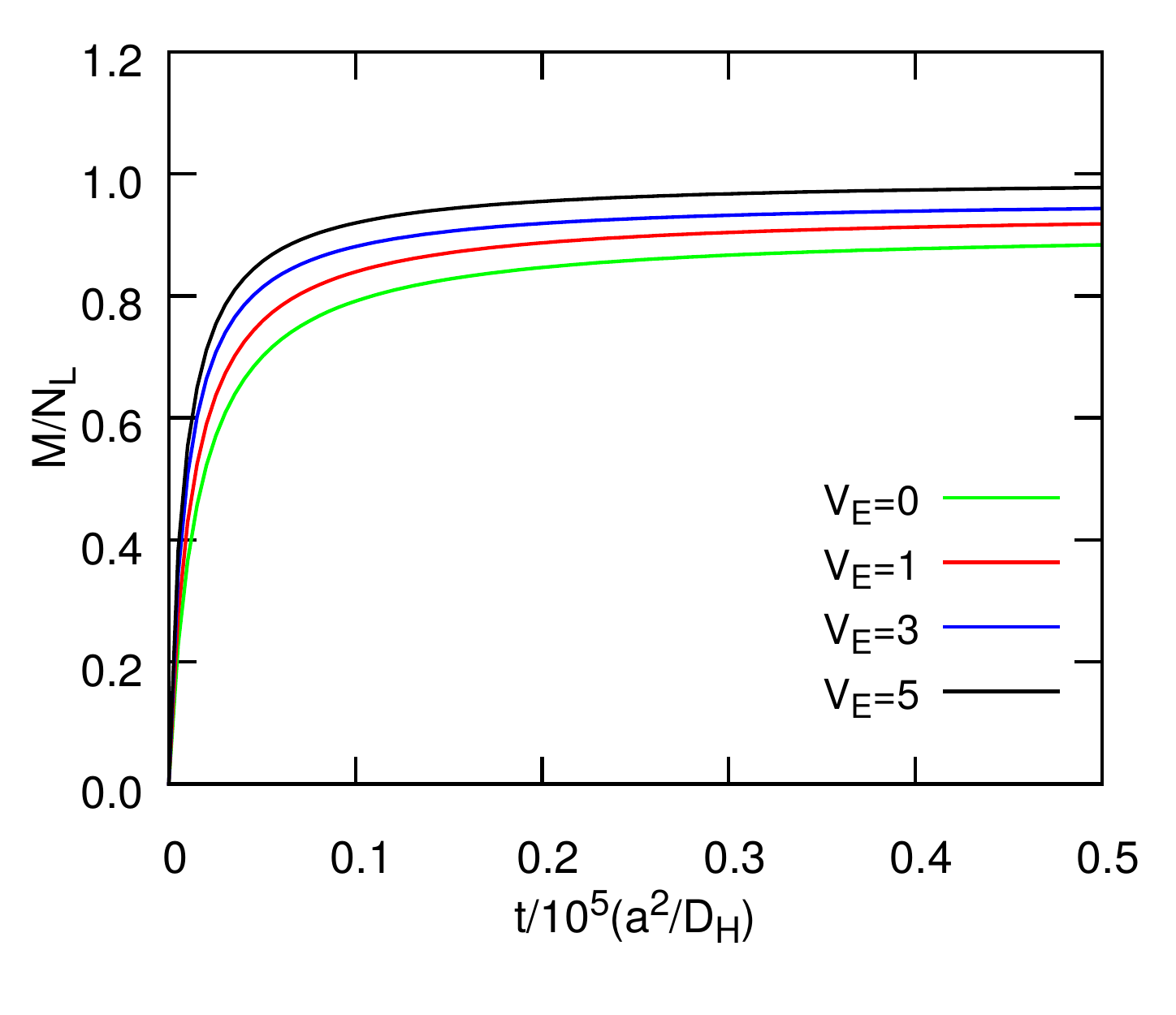}
\caption{Time dependence of the fraction of particles with low diffusion
	coefficient that belong to the largest cluster of this species,
	$M/N_{L}$, for different values of the amplitude of the external
	potential $V_E$, obtained from simulations of $N = 1592$ particles in a
	box of lateral dimension $L_b=100$ ($\rho=0.5$) and $D = 10^{-3}$.
Averages are over $25$ samples.\label{fig::timeclustering}}
\end{figure}
\begin{figure}
\includegraphics[width=\columnwidth]{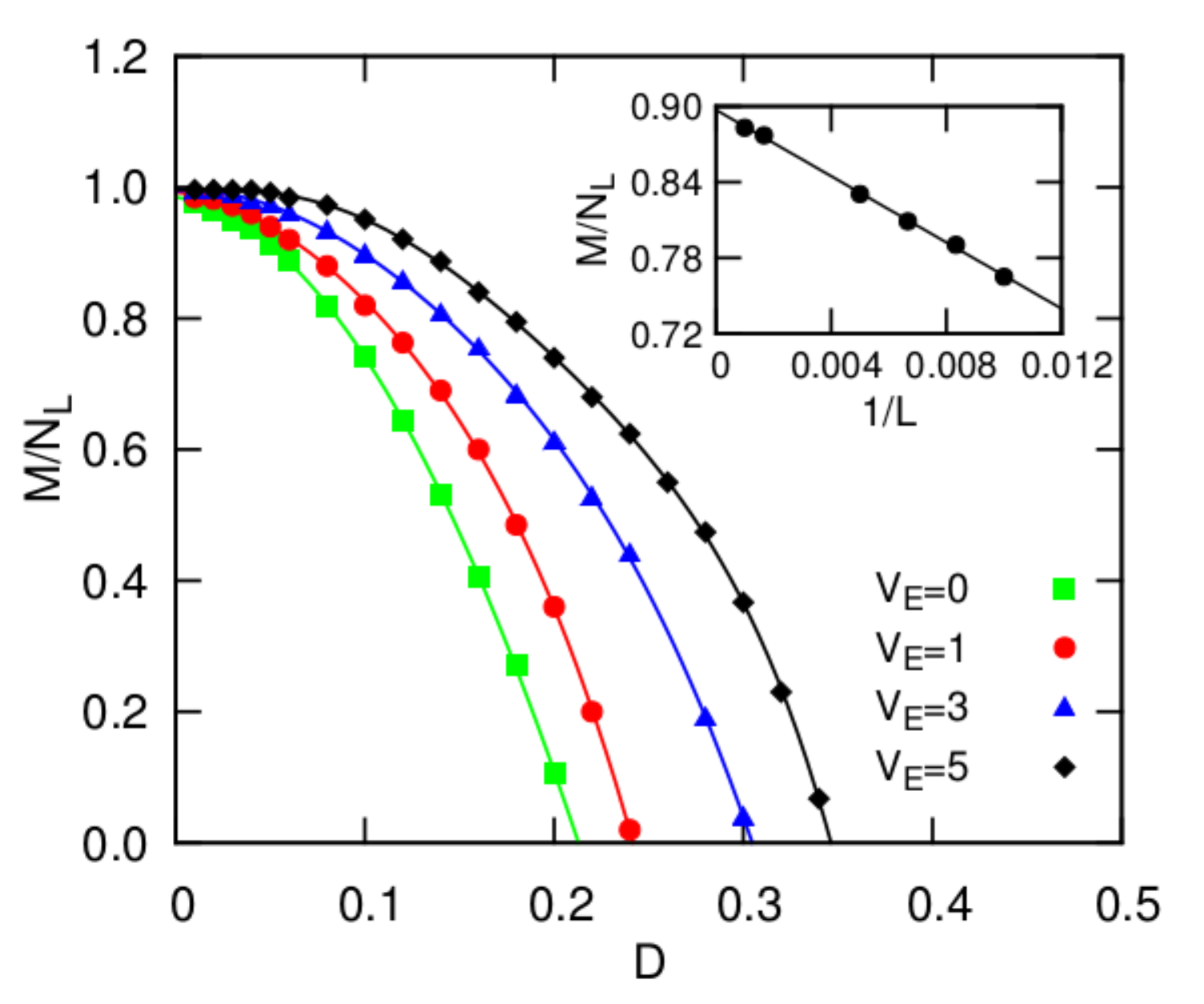}
\caption{Fraction of the particles with low diffusion coefficient that belong
	to the largest cluster of this species, $M/N_{L}$, as a function of the
	ratio $D$. Different symbols correspond to different values of the
	amplitude of the external potential $V_E$, at $\rho=0.5$, obtained by
	extrapolating the dependence of $M/N_{L}$ on the linear size of the box
	$L_b$; boxes with $L_b=\{100,120,150,200,600,1000\}$ were considered.
	The inset shows an example, for $V_E=5$, where the symbols represent
the simulation data and the solid line is the linear
fit.~\label{fig::clustering}}
\end{figure} 

A more quantitative description of the demixing is obtained by calculating the
number $M$ of particles of the largest cluster of the species with the lower
diffusion coefficient ($D_\mathrm{L}$). For the range of parameters where a
band is formed in the center of the box, this cluster corresponds to the band
itself. To identify the cluster, we used Voronoi tessellation~\cite{fortune}.
We divided the space into $N$ cells limited by polygons (one per particle), so
that all the points in a cell are closer to the center of that particle than to
the center of any other particle. Two particles are connected, if they are of
the same species and the corresponding cells are first neighbors in the Voronoi
tessellation. A cluster is then the set of all the connected particles.

Figure~\ref{fig::timeclustering} illustrates the time evolution of the ratio
$M/N_\mathrm{L}$, for different values of the amplitude of the external
potential $V_E$. The value of this ratio saturates asymptotically at a non-zero
value that increases with $V_E$. This is consistent with demixing and cluster
of particles with low diffusion coefficient, as shown in
Fig.~\ref{fig::dandve}. Note that, for $V_E=0$ a non-zero value is also
observed but in this case the largest cluster is not necessarily at the center
of the box.  The dependence of this ratio on $D$ is plotted in
Fig.~\ref{fig::clustering}, where different curves correspond to different
values of $V_E$.  The data points are extrapolations for the values in the
thermodynamic limit, obtained from the dependence of this ratio on the box
size, as shown in the inset.  The lower the value of $D$ the larger the
fraction of particles in the largest cluster. In the limit of vanishing $D$
(i.e., $D_\mathrm{L}\ll D_\mathrm{H}$), the ratio $M/N_\mathrm{L}$ converges to
one. The threshold value of $D$ below which demixing is observed is strongly
affected by the value of $V_E$. In the presence of a sinusoidal external
potential, this threshold increases with $V_{E}$.

\section{Final Remarks\label{sec:4}}
We have shown, using Brownian dynamics simulations, that demixing of binary
mixtures of particles with different effective diffusion coefficients is
significantly enhanced in the presence of an external sinusoidal potential. For
a one-dimensional spatially varying ($x$-dependent) potential with a minimum
along the perpendicular ($y$-) direction, a band of particles with the lowest
diffusion coefficient is formed along the minimum surrounded by particles with
the highest diffusion coefficient. The stronger the external potential the
smaller the difference required between the diffusion coefficients to observe
demixing.

Electromagnetic fields, concentration gradients, and patterned substrates are
examples of external stimuli that have been considered to control the dynamics
of single-component active systems~\cite{erik16,boy16,pince16,mcdermott16}.
Our results suggest that the dynamics of multicomponent systems is much richer
than their single-component counterparts. In particular, demixing may occur and
the conditions required are more general than in the absence of external
stimuli. Future studies may consider the response to other external stimuli and
how differences in the responsiveness of different species might affect the
overall dynamics. Also, inspired by previous results for passive
systems~\cite{araujo2017}, it may be interesting to explore the emergence of
new spatial-temporal patterns in the presence of time-varying stimuli.

\acknowledgments{We acknowledge financial support from the Portuguese
Foundation for Science and Technology(FCT) under Contracts no
UID/FIS/00618/2013 and SFRH/BD/119240/2016.}

\bibliography{my.bib}

\providecommand{\noopsort}[1]{}\providecommand{\singleletter}[1]{#1}%
\begin{thebibliography}{38}%
\makeatletter
\providecommand \@ifxundefined [1]{%
 \@ifx{#1\undefined}
}%
\providecommand \@ifnum [1]{%
 \ifnum #1\expandafter \@firstoftwo
 \else \expandafter \@secondoftwo
 \fi
}%
\providecommand \@ifx [1]{%
 \ifx #1\expandafter \@firstoftwo
 \else \expandafter \@secondoftwo
 \fi
}%
\providecommand \natexlab [1]{#1}%
\providecommand \enquote  [1]{``#1''}%
\providecommand \bibnamefont  [1]{#1}%
\providecommand \bibfnamefont [1]{#1}%
\providecommand \citenamefont [1]{#1}%
\providecommand \href@noop [0]{\@secondoftwo}%
\providecommand \href [0]{\begingroup \@sanitize@url \@href}%
\providecommand \@href[1]{\@@startlink{#1}\@@href}%
\providecommand \@@href[1]{\endgroup#1\@@endlink}%
\providecommand \@sanitize@url [0]{\catcode `\\12\catcode `\$12\catcode
  `\&12\catcode `\#12\catcode `\^12\catcode `\_12\catcode `\%12\relax}%
\providecommand \@@startlink[1]{}%
\providecommand \@@endlink[0]{}%
\providecommand \url  [0]{\begingroup\@sanitize@url \@url }%
\providecommand \@url [1]{\endgroup\@href {#1}{\urlprefix }}%
\providecommand \urlprefix  [0]{URL }%
\providecommand \Eprint [0]{\href }%
\providecommand \doibase [0]{http://dx.doi.org/}%
\providecommand \selectlanguage [0]{\@gobble}%
\providecommand \bibinfo  [0]{\@secondoftwo}%
\providecommand \bibfield  [0]{\@secondoftwo}%
\providecommand \translation [1]{[#1]}%
\providecommand \BibitemOpen [0]{}%
\providecommand \bibitemStop [0]{}%
\providecommand \bibitemNoStop [0]{.\EOS\space}%
\providecommand \EOS [0]{\spacefactor3000\relax}%
\providecommand \BibitemShut  [1]{\csname bibitem#1\endcsname}%
\let\auto@bib@innerbib\@empty
\bibitem [{\citenamefont {Ahmad}\ and\ \citenamefont {Smalley}(1973)}]{ahm73}%
  \BibitemOpen
  \bibfield  {author} {\bibinfo {author} {\bibfnamefont {K.}~\bibnamefont
  {Ahmad}}\ and\ \bibinfo {author} {\bibfnamefont {I.~J.}\ \bibnamefont
  {Smalley}},\ }\href@noop {} {\bibfield  {journal} {\bibinfo  {journal}
  {Powder Technol.}\ }\textbf {\bibinfo {volume} {9}},\ \bibinfo {pages} {69}
  (\bibinfo {year} {1973})}\BibitemShut {NoStop}%
\bibitem [{\citenamefont {Dzubiella}\ and\ \citenamefont
  {L{\"o}wen}(2002)}]{dzu02}%
  \BibitemOpen
  \bibfield  {author} {\bibinfo {author} {\bibfnamefont {J.}~\bibnamefont
  {Dzubiella}}\ and\ \bibinfo {author} {\bibfnamefont {H.}~\bibnamefont
  {L{\"o}wen}},\ }\href@noop {} {\bibfield  {journal} {\bibinfo  {journal} {J.
  Phys. Condens. Matt}\ }\textbf {\bibinfo {volume} {14}},\ \bibinfo {pages}
  {9383} (\bibinfo {year} {2002})}\BibitemShut {NoStop}%
\bibitem [{\citenamefont {Kudrolli}(2004)}]{kud04}%
  \BibitemOpen
  \bibfield  {author} {\bibinfo {author} {\bibfnamefont {A.}~\bibnamefont
  {Kudrolli}},\ }\href@noop {} {\bibfield  {journal} {\bibinfo  {journal} {Rep.
  Prog. Phys.}\ }\textbf {\bibinfo {volume} {67}},\ \bibinfo {pages} {209}
  (\bibinfo {year} {2004})}\BibitemShut {NoStop}%
\bibitem [{\citenamefont {Leunissen}\ \emph {et~al.}(2005)\citenamefont
  {Leunissen}, \citenamefont {Christova}, \citenamefont {Hynninen},
  \citenamefont {Royall}, \citenamefont {Campbell}, \citenamefont {Imhof},
  \citenamefont {Dijkstra}, \citenamefont {van Roij},\ and\ \citenamefont {van
  Blaaderen}}]{leu05}%
  \BibitemOpen
  \bibfield  {author} {\bibinfo {author} {\bibfnamefont {M.~E.}\ \bibnamefont
  {Leunissen}}, \bibinfo {author} {\bibfnamefont {C.~G.}\ \bibnamefont
  {Christova}}, \bibinfo {author} {\bibfnamefont {A.~P.}\ \bibnamefont
  {Hynninen}}, \bibinfo {author} {\bibfnamefont {C.~P.}\ \bibnamefont
  {Royall}}, \bibinfo {author} {\bibfnamefont {A.~I.}\ \bibnamefont
  {Campbell}}, \bibinfo {author} {\bibfnamefont {A.}~\bibnamefont {Imhof}},
  \bibinfo {author} {\bibfnamefont {M.}~\bibnamefont {Dijkstra}}, \bibinfo
  {author} {\bibfnamefont {R.}~\bibnamefont {van Roij}}, \ and\ \bibinfo
  {author} {\bibfnamefont {A.}~\bibnamefont {van Blaaderen}},\ }\href@noop {}
  {\bibfield  {journal} {\bibinfo  {journal} {Nature}\ }\textbf {\bibinfo
  {volume} {437}},\ \bibinfo {pages} {235} (\bibinfo {year}
  {2005})}\BibitemShut {NoStop}%
\bibitem [{\citenamefont {Ehrhardt}\ \emph {et~al.}(2005)\citenamefont
  {Ehrhardt}, \citenamefont {Stephenson},\ and\ \citenamefont {Reis}}]{gcma05}%
  \BibitemOpen
  \bibfield  {author} {\bibinfo {author} {\bibfnamefont {G.~C. M.~A.}\
  \bibnamefont {Ehrhardt}}, \bibinfo {author} {\bibfnamefont {A.}~\bibnamefont
  {Stephenson}}, \ and\ \bibinfo {author} {\bibfnamefont {P.~M.}\ \bibnamefont
  {Reis}},\ }\href@noop {} {\bibfield  {journal} {\bibinfo  {journal} {Phys.
  Rev. E.}\ }\textbf {\bibinfo {volume} {71}},\ \bibinfo {pages} {041301}
  (\bibinfo {year} {2005})}\BibitemShut {NoStop}%
\bibitem [{\citenamefont {Reis}\ \emph {et~al.}(2006)\citenamefont {Reis},
  \citenamefont {Sykes},\ and\ \citenamefont {Mullin}}]{reis06}%
  \BibitemOpen
  \bibfield  {author} {\bibinfo {author} {\bibfnamefont {P.~M.}\ \bibnamefont
  {Reis}}, \bibinfo {author} {\bibfnamefont {T.}~\bibnamefont {Sykes}}, \ and\
  \bibinfo {author} {\bibfnamefont {T.}~\bibnamefont {Mullin}},\ }\href@noop {}
  {\bibfield  {journal} {\bibinfo  {journal} {Phys. Rev. E.}\ }\textbf
  {\bibinfo {volume} {74}},\ \bibinfo {pages} {051306} (\bibinfo {year}
  {2006})}\BibitemShut {NoStop}%
\bibitem [{\citenamefont {Vissers}\ \emph {et~al.}(2011)\citenamefont
  {Vissers}, \citenamefont {van Blaaderen},\ and\ \citenamefont
  {Imhof}}]{vis11}%
  \BibitemOpen
  \bibfield  {author} {\bibinfo {author} {\bibfnamefont {T.}~\bibnamefont
  {Vissers}}, \bibinfo {author} {\bibfnamefont {A.}~\bibnamefont {van
  Blaaderen}}, \ and\ \bibinfo {author} {\bibfnamefont {A.}~\bibnamefont
  {Imhof}},\ }\href@noop {} {\bibfield  {journal} {\bibinfo  {journal} {Phys.
  Rev. Lett.}\ }\textbf {\bibinfo {volume} {106}},\ \bibinfo {pages} {228303}
  (\bibinfo {year} {2011})}\BibitemShut {NoStop}%
\bibitem [{\citenamefont {Rivas}\ \emph
  {et~al.}(2011{\natexlab{a}})\citenamefont {Rivas}, \citenamefont {Cordero},
  \citenamefont {Risso},\ and\ \citenamefont {Soto}}]{riv11}%
  \BibitemOpen
  \bibfield  {author} {\bibinfo {author} {\bibfnamefont {N.}~\bibnamefont
  {Rivas}}, \bibinfo {author} {\bibfnamefont {P.}~\bibnamefont {Cordero}},
  \bibinfo {author} {\bibfnamefont {D.}~\bibnamefont {Risso}}, \ and\ \bibinfo
  {author} {\bibfnamefont {R.}~\bibnamefont {Soto}},\ }\href@noop {} {\bibfield
   {journal} {\bibinfo  {journal} {New. J. Phys.}\ }\textbf {\bibinfo {volume}
  {13}},\ \bibinfo {pages} {055018} (\bibinfo {year}
  {2011}{\natexlab{a}})}\BibitemShut {NoStop}%
\bibitem [{\citenamefont {Rivas}\ \emph {et~al.}(2012)\citenamefont {Rivas},
  \citenamefont {Cordero}, \citenamefont {Risso},\ and\ \citenamefont
  {Soto}}]{riv12}%
  \BibitemOpen
  \bibfield  {author} {\bibinfo {author} {\bibfnamefont {N.}~\bibnamefont
  {Rivas}}, \bibinfo {author} {\bibfnamefont {P.}~\bibnamefont {Cordero}},
  \bibinfo {author} {\bibfnamefont {D.}~\bibnamefont {Risso}}, \ and\ \bibinfo
  {author} {\bibfnamefont {R.}~\bibnamefont {Soto}},\ }\href@noop {} {\bibfield
   {journal} {\bibinfo  {journal} {Granular Matter}\ }\textbf {\bibinfo
  {volume} {14}},\ \bibinfo {pages} {157} (\bibinfo {year} {2012})}\BibitemShut
  {NoStop}%
\bibitem [{\citenamefont {Gr{\"u}nwald}\ \emph {et~al.}(2016)\citenamefont
  {Gr{\"u}nwald}, \citenamefont {Tricard}, \citenamefont {Whitesides},\ and\
  \citenamefont {Geissler}}]{grun16}%
  \BibitemOpen
  \bibfield  {author} {\bibinfo {author} {\bibfnamefont {M.}~\bibnamefont
  {Gr{\"u}nwald}}, \bibinfo {author} {\bibfnamefont {S.}~\bibnamefont
  {Tricard}}, \bibinfo {author} {\bibfnamefont {G.~M.}\ \bibnamefont
  {Whitesides}}, \ and\ \bibinfo {author} {\bibfnamefont {P.~L.}\ \bibnamefont
  {Geissler}},\ }\href@noop {} {\bibfield  {journal} {\bibinfo  {journal} {Soft
  Matter}\ }\textbf {\bibinfo {volume} {12}},\ \bibinfo {pages} {1517}
  (\bibinfo {year} {2016})}\BibitemShut {NoStop}%
\bibitem [{\citenamefont {Rivas}\ \emph
  {et~al.}(2011{\natexlab{b}})\citenamefont {Rivas}, \citenamefont {Ponce},
  \citenamefont {Gallet}, \citenamefont {Risso}, \citenamefont {Soto},
  \citenamefont {Cordero},\ and\ \citenamefont {Mujica}}]{riva11}%
  \BibitemOpen
  \bibfield  {author} {\bibinfo {author} {\bibfnamefont {N.}~\bibnamefont
  {Rivas}}, \bibinfo {author} {\bibfnamefont {S.}~\bibnamefont {Ponce}},
  \bibinfo {author} {\bibfnamefont {B.}~\bibnamefont {Gallet}}, \bibinfo
  {author} {\bibfnamefont {D.}~\bibnamefont {Risso}}, \bibinfo {author}
  {\bibfnamefont {R.}~\bibnamefont {Soto}}, \bibinfo {author} {\bibfnamefont
  {P.}~\bibnamefont {Cordero}}, \ and\ \bibinfo {author} {\bibfnamefont
  {N.}~\bibnamefont {Mujica}},\ }\href@noop {} {\bibfield  {journal} {\bibinfo
  {journal} {Phys. Rev. Lett.}\ }\textbf {\bibinfo {volume} {106}},\ \bibinfo
  {pages} {088001} (\bibinfo {year} {2011}{\natexlab{b}})}\BibitemShut
  {NoStop}%
\bibitem [{\citenamefont {Stenhammar}\ \emph {et~al.}(2015)\citenamefont
  {Stenhammar}, \citenamefont {Wittkowski}, \citenamefont {Marenduzzo},\ and\
  \citenamefont {Cates}}]{sten15}%
  \BibitemOpen
  \bibfield  {author} {\bibinfo {author} {\bibfnamefont {J.}~\bibnamefont
  {Stenhammar}}, \bibinfo {author} {\bibfnamefont {R.}~\bibnamefont
  {Wittkowski}}, \bibinfo {author} {\bibfnamefont {D.}~\bibnamefont
  {Marenduzzo}}, \ and\ \bibinfo {author} {\bibfnamefont {M.~E.}\ \bibnamefont
  {Cates}},\ }\href@noop {} {\bibfield  {journal} {\bibinfo  {journal} {Phys.
  Rev. Lett.}\ }\textbf {\bibinfo {volume} {114}},\ \bibinfo {pages} {018301}
  (\bibinfo {year} {2015})}\BibitemShut {NoStop}%
\bibitem [{\citenamefont {Weber}\ \emph {et~al.}(2016)\citenamefont {Weber},
  \citenamefont {Weber},\ and\ \citenamefont {Frey}}]{sim16}%
  \BibitemOpen
  \bibfield  {author} {\bibinfo {author} {\bibfnamefont {S.~N.}\ \bibnamefont
  {Weber}}, \bibinfo {author} {\bibfnamefont {C.~A.}\ \bibnamefont {Weber}}, \
  and\ \bibinfo {author} {\bibfnamefont {E.}~\bibnamefont {Frey}},\ }\href@noop
  {} {\bibfield  {journal} {\bibinfo  {journal} {Phys. Rev. Lett.}\ }\textbf
  {\bibinfo {volume} {116}},\ \bibinfo {pages} {058301} (\bibinfo {year}
  {2016})}\BibitemShut {NoStop}%
\bibitem [{\citenamefont {Ramaswamy}(2010)}]{Ram10}%
  \BibitemOpen
  \bibfield  {author} {\bibinfo {author} {\bibfnamefont {S.}~\bibnamefont
  {Ramaswamy}},\ }\href@noop {} {\bibfield  {journal} {\bibinfo  {journal}
  {Annu. Rev. Condens. Matt. Phys.}\ }\textbf {\bibinfo {volume} {1}},\
  \bibinfo {pages} {323} (\bibinfo {year} {2010})}\BibitemShut {NoStop}%
\bibitem [{\citenamefont {Bechinger}\ \emph {et~al.}(2016)\citenamefont
  {Bechinger}, \citenamefont {Leonardo}, \citenamefont {L{\"o}wen},
  \citenamefont {Reichhardt}, \citenamefont {Volpe},\ and\ \citenamefont
  {Volpe}}]{bech16}%
  \BibitemOpen
  \bibfield  {author} {\bibinfo {author} {\bibfnamefont {C.}~\bibnamefont
  {Bechinger}}, \bibinfo {author} {\bibfnamefont {R.~D.}\ \bibnamefont
  {Leonardo}}, \bibinfo {author} {\bibfnamefont {H.}~\bibnamefont {L{\"o}wen}},
  \bibinfo {author} {\bibfnamefont {C.}~\bibnamefont {Reichhardt}}, \bibinfo
  {author} {\bibfnamefont {G.}~\bibnamefont {Volpe}}, \ and\ \bibinfo {author}
  {\bibfnamefont {G.}~\bibnamefont {Volpe}},\ }\href@noop {} {\bibfield
  {journal} {\bibinfo  {journal} {Rev. Mod. Phys.}\ }\textbf {\bibinfo {volume}
  {88}},\ \bibinfo {pages} {045006} (\bibinfo {year} {2016})}\BibitemShut
  {NoStop}%
\bibitem [{\citenamefont {Trepat}\ \emph {et~al.}(2009)\citenamefont {Trepat},
  \citenamefont {Wasserman}, \citenamefont {Angelini}, \citenamefont
  {E.~Millet}, \citenamefont {Butler},\ and\ \citenamefont {Fredberg}}]{tre09}%
  \BibitemOpen
  \bibfield  {author} {\bibinfo {author} {\bibfnamefont {X.}~\bibnamefont
  {Trepat}}, \bibinfo {author} {\bibfnamefont {M.~R.}\ \bibnamefont
  {Wasserman}}, \bibinfo {author} {\bibfnamefont {T.~E.}\ \bibnamefont
  {Angelini}}, \bibinfo {author} {\bibfnamefont {D.~A.~W.}\ \bibnamefont
  {E.~Millet}}, \bibinfo {author} {\bibfnamefont {J.~P.}\ \bibnamefont
  {Butler}}, \ and\ \bibinfo {author} {\bibfnamefont {J.~J.}\ \bibnamefont
  {Fredberg}},\ }\href@noop {} {\bibfield  {journal} {\bibinfo  {journal}
  {Nature Physics}\ }\textbf {\bibinfo {volume} {5}},\ \bibinfo {pages} {426}
  (\bibinfo {year} {2009})}\BibitemShut {NoStop}%
\bibitem [{\citenamefont {Angelini}\ \emph {et~al.}(2011)\citenamefont
  {Angelini}, \citenamefont {Hannezo}, \citenamefont {Trepat}, \citenamefont
  {Marquez}, \citenamefont {Fredberg},\ and\ \citenamefont {Weitz}}]{ang11}%
  \BibitemOpen
  \bibfield  {author} {\bibinfo {author} {\bibfnamefont {T.~E.}\ \bibnamefont
  {Angelini}}, \bibinfo {author} {\bibfnamefont {E.}~\bibnamefont {Hannezo}},
  \bibinfo {author} {\bibfnamefont {X.}~\bibnamefont {Trepat}}, \bibinfo
  {author} {\bibfnamefont {M.}~\bibnamefont {Marquez}}, \bibinfo {author}
  {\bibfnamefont {J.~J.}\ \bibnamefont {Fredberg}}, \ and\ \bibinfo {author}
  {\bibfnamefont {D.~A.}\ \bibnamefont {Weitz}},\ }\href@noop {} {\bibfield
  {journal} {\bibinfo  {journal} {PNAS}\ }\textbf {\bibinfo {volume} {108}},\
  \bibinfo {pages} {4714} (\bibinfo {year} {2011})}\BibitemShut {NoStop}%
\bibitem [{\citenamefont {Chai}\ \emph {et~al.}(2011)\citenamefont {Chai},
  \citenamefont {Vlamakis},\ and\ \citenamefont {Kolter}}]{chai11}%
  \BibitemOpen
  \bibfield  {author} {\bibinfo {author} {\bibfnamefont {L.}~\bibnamefont
  {Chai}}, \bibinfo {author} {\bibfnamefont {H.}~\bibnamefont {Vlamakis}}, \
  and\ \bibinfo {author} {\bibfnamefont {R.}~\bibnamefont {Kolter}},\
  }\href@noop {} {\bibfield  {journal} {\bibinfo  {journal} {MRS Bulletin}\
  }\textbf {\bibinfo {volume} {36}},\ \bibinfo {pages} {374} (\bibinfo {year}
  {2011})}\BibitemShut {NoStop}%
\bibitem [{\citenamefont {Drescher}\ \emph {et~al.}(2011)\citenamefont
  {Drescher}, \citenamefont {Dunkel}, \citenamefont {Cisneros}, \citenamefont
  {Ganguly},\ and\ \citenamefont {Goldstein}}]{dres11}%
  \BibitemOpen
  \bibfield  {author} {\bibinfo {author} {\bibfnamefont {K.}~\bibnamefont
  {Drescher}}, \bibinfo {author} {\bibfnamefont {J.}~\bibnamefont {Dunkel}},
  \bibinfo {author} {\bibfnamefont {L.~H.}\ \bibnamefont {Cisneros}}, \bibinfo
  {author} {\bibfnamefont {S.}~\bibnamefont {Ganguly}}, \ and\ \bibinfo
  {author} {\bibfnamefont {R.~E.}\ \bibnamefont {Goldstein}},\ }\href@noop {}
  {\bibfield  {journal} {\bibinfo  {journal} {PNAS}\ }\textbf {\bibinfo
  {volume} {108}},\ \bibinfo {pages} {10940} (\bibinfo {year}
  {2011})}\BibitemShut {NoStop}%
\bibitem [{\citenamefont {Ni}\ \emph {et~al.}(2014)\citenamefont {Ni},
  \citenamefont {Stuart}, \citenamefont {Dijkstra},\ and\ \citenamefont
  {Bolhuis}}]{nir14}%
  \BibitemOpen
  \bibfield  {author} {\bibinfo {author} {\bibfnamefont {R.}~\bibnamefont
  {Ni}}, \bibinfo {author} {\bibfnamefont {M.~A.~C.}\ \bibnamefont {Stuart}},
  \bibinfo {author} {\bibfnamefont {M.}~\bibnamefont {Dijkstra}}, \ and\
  \bibinfo {author} {\bibfnamefont {P.~G.}\ \bibnamefont {Bolhuis}},\
  }\href@noop {} {\bibfield  {journal} {\bibinfo  {journal} {Soft Matter}\
  }\textbf {\bibinfo {volume} {10}},\ \bibinfo {pages} {6609} (\bibinfo {year}
  {2014})}\BibitemShut {NoStop}%
\bibitem [{\citenamefont {Kummel}\ \emph {et~al.}(2015)\citenamefont {Kummel},
  \citenamefont {Shabestari}, \citenamefont {Lozano}, \citenamefont {Volpe},\
  and\ \citenamefont {Bechinger}}]{kum15}%
  \BibitemOpen
  \bibfield  {author} {\bibinfo {author} {\bibfnamefont {F.}~\bibnamefont
  {Kummel}}, \bibinfo {author} {\bibfnamefont {P.}~\bibnamefont {Shabestari}},
  \bibinfo {author} {\bibfnamefont {C.}~\bibnamefont {Lozano}}, \bibinfo
  {author} {\bibfnamefont {G.}~\bibnamefont {Volpe}}, \ and\ \bibinfo {author}
  {\bibfnamefont {C.}~\bibnamefont {Bechinger}},\ }\href@noop {} {\bibfield
  {journal} {\bibinfo  {journal} {Soft Matter}\ }\textbf {\bibinfo {volume}
  {11}},\ \bibinfo {pages} {6187} (\bibinfo {year} {2015})}\BibitemShut
  {NoStop}%
\bibitem [{\citenamefont {Takatori}\ and\ \citenamefont
  {Brady}(2015)}]{taka15}%
  \BibitemOpen
  \bibfield  {author} {\bibinfo {author} {\bibfnamefont {S.~C.}\ \bibnamefont
  {Takatori}}\ and\ \bibinfo {author} {\bibfnamefont {J.~F.}\ \bibnamefont
  {Brady}},\ }\href@noop {} {\bibfield  {journal} {\bibinfo  {journal} {Soft
  Matter}\ }\textbf {\bibinfo {volume} {11}},\ \bibinfo {pages} {7920}
  (\bibinfo {year} {2015})}\BibitemShut {NoStop}%
\bibitem [{\citenamefont {Smrek}\ and\ \citenamefont {Kremer}(2017)}]{sme17}%
  \BibitemOpen
  \bibfield  {author} {\bibinfo {author} {\bibfnamefont {J.}~\bibnamefont
  {Smrek}}\ and\ \bibinfo {author} {\bibfnamefont {K.}~\bibnamefont {Kremer}},\
  }\href@noop {} {\bibfield  {journal} {\bibinfo  {journal} {Phys. Rev. Lett.}\
  }\textbf {\bibinfo {volume} {118}},\ \bibinfo {pages} {098002} (\bibinfo
  {year} {2017})}\BibitemShut {NoStop}%
\bibitem [{\citenamefont {Yang}\ \emph {et~al.}(2014)\citenamefont {Yang},
  \citenamefont {Manning},\ and\ \citenamefont {Marchetti}}]{yang14}%
  \BibitemOpen
  \bibfield  {author} {\bibinfo {author} {\bibfnamefont {X.}~\bibnamefont
  {Yang}}, \bibinfo {author} {\bibfnamefont {M.~L.}\ \bibnamefont {Manning}}, \
  and\ \bibinfo {author} {\bibfnamefont {M.~C.}\ \bibnamefont {Marchetti}},\
  }\href@noop {} {\bibfield  {journal} {\bibinfo  {journal} {Soft Matter}\
  }\textbf {\bibinfo {volume} {10}},\ \bibinfo {pages} {6477} (\bibinfo {year}
  {2014})}\BibitemShut {NoStop}%
\bibitem [{\citenamefont {Grosberg}\ and\ \citenamefont
  {Joanny}(2015)}]{gros15}%
  \BibitemOpen
  \bibfield  {author} {\bibinfo {author} {\bibfnamefont {A.~Y.}\ \bibnamefont
  {Grosberg}}\ and\ \bibinfo {author} {\bibfnamefont {J.~F.}\ \bibnamefont
  {Joanny}},\ }\href@noop {} {\bibfield  {journal} {\bibinfo  {journal} {Phys.
  Rev. E.}\ }\textbf {\bibinfo {volume} {92}},\ \bibinfo {pages} {032118}
  (\bibinfo {year} {2015})}\BibitemShut {NoStop}%
\bibitem [{\citenamefont {Tanaka}\ \emph {et~al.}(2017)\citenamefont {Tanaka},
  \citenamefont {Lee},\ and\ \citenamefont {Brenner}}]{tana17}%
  \BibitemOpen
  \bibfield  {author} {\bibinfo {author} {\bibfnamefont {H.}~\bibnamefont
  {Tanaka}}, \bibinfo {author} {\bibfnamefont {A.~A.}\ \bibnamefont {Lee}}, \
  and\ \bibinfo {author} {\bibfnamefont {M.~P.}\ \bibnamefont {Brenner}},\
  }\href@noop {} {\bibfield  {journal} {\bibinfo  {journal} {Phys. Rev.
  Fluids}\ }\textbf {\bibinfo {volume} {2}},\ \bibinfo {pages} {043103}
  (\bibinfo {year} {2017})}\BibitemShut {NoStop}%
\bibitem [{\citenamefont {Adler}(1966)}]{adl66}%
  \BibitemOpen
  \bibfield  {author} {\bibinfo {author} {\bibfnamefont {J.}~\bibnamefont
  {Adler}},\ }\href@noop {} {\bibfield  {journal} {\bibinfo  {journal}
  {Science}\ }\textbf {\bibinfo {volume} {153}},\ \bibinfo {pages} {708}
  (\bibinfo {year} {1966})}\BibitemShut {NoStop}%
\bibitem [{\citenamefont {Armitage}\ and\ \citenamefont
  {Hellingwerf}(2003)}]{arm03}%
  \BibitemOpen
  \bibfield  {author} {\bibinfo {author} {\bibfnamefont {J.~P.}\ \bibnamefont
  {Armitage}}\ and\ \bibinfo {author} {\bibfnamefont {K.~J.}\ \bibnamefont
  {Hellingwerf}},\ }\href@noop {} {\bibfield  {journal} {\bibinfo  {journal}
  {Photosynth. Res}\ }\textbf {\bibinfo {volume} {76}},\ \bibinfo {pages} {145}
  (\bibinfo {year} {2003})}\BibitemShut {NoStop}%
\bibitem [{\citenamefont {Blakemore}(1975)}]{blak75}%
  \BibitemOpen
  \bibfield  {author} {\bibinfo {author} {\bibfnamefont {R.}~\bibnamefont
  {Blakemore}},\ }\href@noop {} {\bibfield  {journal} {\bibinfo  {journal}
  {Science}\ }\textbf {\bibinfo {volume} {190}},\ \bibinfo {pages} {377}
  (\bibinfo {year} {1975})}\BibitemShut {NoStop}%
\bibitem [{\citenamefont {Bricard}\ \emph {et~al.}(2013)\citenamefont
  {Bricard}, \citenamefont {Caussin}, \citenamefont {Desreumaux}, \citenamefont
  {Dauchot},\ and\ \citenamefont {Bartolo}}]{ant13}%
  \BibitemOpen
  \bibfield  {author} {\bibinfo {author} {\bibfnamefont {A.}~\bibnamefont
  {Bricard}}, \bibinfo {author} {\bibfnamefont {J.~B.}\ \bibnamefont
  {Caussin}}, \bibinfo {author} {\bibfnamefont {N.}~\bibnamefont {Desreumaux}},
  \bibinfo {author} {\bibfnamefont {O.}~\bibnamefont {Dauchot}}, \ and\
  \bibinfo {author} {\bibfnamefont {D.}~\bibnamefont {Bartolo}},\ }\href@noop
  {} {\bibfield  {journal} {\bibinfo  {journal} {Nature}\ }\textbf {\bibinfo
  {volume} {503}},\ \bibinfo {pages} {95} (\bibinfo {year} {2013})}\BibitemShut
  {NoStop}%
\bibitem [{\citenamefont {Yan}\ \emph {et~al.}(2016)\citenamefont {Yan},
  \citenamefont {Han}, \citenamefont {Zhang}, \citenamefont {Xu}, \citenamefont
  {Lujiten},\ and\ \citenamefont {Granick}}]{erik16}%
  \BibitemOpen
  \bibfield  {author} {\bibinfo {author} {\bibfnamefont {J.}~\bibnamefont
  {Yan}}, \bibinfo {author} {\bibfnamefont {M.}~\bibnamefont {Han}}, \bibinfo
  {author} {\bibfnamefont {J.}~\bibnamefont {Zhang}}, \bibinfo {author}
  {\bibfnamefont {C.}~\bibnamefont {Xu}}, \bibinfo {author} {\bibfnamefont
  {E.}~\bibnamefont {Lujiten}}, \ and\ \bibinfo {author} {\bibfnamefont
  {S.}~\bibnamefont {Granick}},\ }\href@noop {} {\bibfield  {journal} {\bibinfo
   {journal} {Nature Materials}\ }\textbf {\bibinfo {volume} {15}},\ \bibinfo
  {pages} {1095} (\bibinfo {year} {2016})}\BibitemShut {NoStop}%
\bibitem [{\citenamefont {Boymelgreen}\ \emph {et~al.}(2016)\citenamefont
  {Boymelgreen}, \citenamefont {Yossifon},\ and\ \citenamefont
  {Miloh}}]{boy16}%
  \BibitemOpen
  \bibfield  {author} {\bibinfo {author} {\bibfnamefont {A.}~\bibnamefont
  {Boymelgreen}}, \bibinfo {author} {\bibfnamefont {G.}~\bibnamefont
  {Yossifon}}, \ and\ \bibinfo {author} {\bibfnamefont {T.}~\bibnamefont
  {Miloh}},\ }\href@noop {} {\bibfield  {journal} {\bibinfo  {journal}
  {Langmuir}\ }\textbf {\bibinfo {volume} {32}},\ \bibinfo {pages} {9540}
  (\bibinfo {year} {2016})}\BibitemShut {NoStop}%
\bibitem [{\citenamefont {Branka}\ and\ \citenamefont {Heyes}(1999)}]{Branka}%
  \BibitemOpen
  \bibfield  {author} {\bibinfo {author} {\bibfnamefont {A.~C.}\ \bibnamefont
  {Branka}}\ and\ \bibinfo {author} {\bibfnamefont {D.~M.}\ \bibnamefont
  {Heyes}},\ }\href@noop {} {\bibfield  {journal} {\bibinfo  {journal} {Phys.
  Rev. E}\ }\textbf {\bibinfo {volume} {60}},\ \bibinfo {pages} {2381}
  (\bibinfo {year} {1999})}\BibitemShut {NoStop}%
\bibitem [{\citenamefont {Nunes}\ \emph {et~al.}(2016)\citenamefont {Nunes},
  \citenamefont {Araujo},\ and\ \citenamefont {da. Gama}}]{and16}%
  \BibitemOpen
  \bibfield  {author} {\bibinfo {author} {\bibfnamefont {A.~S.}\ \bibnamefont
  {Nunes}}, \bibinfo {author} {\bibfnamefont {N.~A.~M.}\ \bibnamefont
  {Araujo}}, \ and\ \bibinfo {author} {\bibfnamefont {M.~M.~T.}\ \bibnamefont
  {da. Gama}},\ }\href@noop {} {\bibfield  {journal} {\bibinfo  {journal} {J.
  Chem. Phys.}\ }\textbf {\bibinfo {volume} {44}},\ \bibinfo {pages} {034902}
  (\bibinfo {year} {2016})}\BibitemShut {NoStop}%
\bibitem [{\citenamefont {Fortune}(1987)}]{fortune}%
  \BibitemOpen
  \bibfield  {author} {\bibinfo {author} {\bibfnamefont {S.}~\bibnamefont
  {Fortune}},\ }\href@noop {} {\bibfield  {journal} {\bibinfo  {journal}
  {Algorithmica}\ }\textbf {\bibinfo {volume} {2}},\ \bibinfo {pages} {153}
  (\bibinfo {year} {1987})}\BibitemShut {NoStop}%
\bibitem [{\citenamefont {Pin\c{c}e}\ \emph {et~al.}(2016)\citenamefont
  {Pin\c{c}e}, \citenamefont {Velu}, \citenamefont {Callegari}, \citenamefont
  {Elahi}, \citenamefont {Gigan}, \citenamefont {Volpe},\ and\ \citenamefont
  {Volpe}}]{pince16}%
  \BibitemOpen
  \bibfield  {author} {\bibinfo {author} {\bibfnamefont {E.}~\bibnamefont
  {Pin\c{c}e}}, \bibinfo {author} {\bibfnamefont {S.~K.~P.}\ \bibnamefont
  {Velu}}, \bibinfo {author} {\bibfnamefont {A.}~\bibnamefont {Callegari}},
  \bibinfo {author} {\bibfnamefont {P.}~\bibnamefont {Elahi}}, \bibinfo
  {author} {\bibfnamefont {S.}~\bibnamefont {Gigan}}, \bibinfo {author}
  {\bibfnamefont {G.}~\bibnamefont {Volpe}}, \ and\ \bibinfo {author}
  {\bibfnamefont {G.}~\bibnamefont {Volpe}},\ }\href@noop {} {\bibfield
  {journal} {\bibinfo  {journal} {Nat. Comm.}\ }\textbf {\bibinfo {volume}
  {7}},\ \bibinfo {pages} {10907} (\bibinfo {year} {2016})}\BibitemShut
  {NoStop}%
\bibitem [{\citenamefont {\mbox{McDermott}}\ \emph {et~al.}(2016)\citenamefont
  {\mbox{McDermott}}, \citenamefont {Reichhardt},\ and\ \citenamefont
  {Reichhardt}}]{mcdermott16}%
  \BibitemOpen
  \bibfield  {author} {\bibinfo {author} {\bibfnamefont {D.}~\bibnamefont
  {\mbox{McDermott}}}, \bibinfo {author} {\bibfnamefont {C.~J.~O.}\
  \bibnamefont {Reichhardt}}, \ and\ \bibinfo {author} {\bibfnamefont
  {C.}~\bibnamefont {Reichhardt}},\ }\href@noop {} {\bibfield  {journal}
  {\bibinfo  {journal} {Soft Matter}\ }\textbf {\bibinfo {volume} {12}},\
  \bibinfo {pages} {8606} (\bibinfo {year} {2016})}\BibitemShut {NoStop}%
\bibitem [{\citenamefont {Ara\'ujo}\ \emph {et~al.}()\citenamefont {Ara\'ujo},
  \citenamefont {Zezyulin}, \citenamefont {Konotop},\ and\ \citenamefont
  {da~Gama}}]{araujo2017}%
  \BibitemOpen
  \bibfield  {author} {\bibinfo {author} {\bibfnamefont {N.~A.~M.}\
  \bibnamefont {Ara\'ujo}}, \bibinfo {author} {\bibfnamefont {D.~A.}\
  \bibnamefont {Zezyulin}}, \bibinfo {author} {\bibfnamefont {V.~V.}\
  \bibnamefont {Konotop}}, \ and\ \bibinfo {author} {\bibfnamefont {M.~M.~T.}\
  \bibnamefont {da~Gama}},\ }\href@noop {} {\enquote {\bibinfo {title}
  {Dynamical design of spatial patterns of colloidal suspensions},}\ }\Eprint
  {http://arxiv.org/abs/arXiv:1706.02067} {arXiv:1706.02067} \BibitemShut
  {NoStop}%
\end{thebibliography}%
\end{document}